\documentclass[ajkp,dda,twocolumn,groupedaddress]{revtex4}

\usepackage{graphicx,amssymb}

\begin{document}

\title{Temperature dependance of the tunneling density of states in sub-micron planar metal / oxide / graphene junctions}
\author{ S Hacohen-Gourgy}
\email{hacohe@post.tau.ac.il}
\author{I Diamant}
\author{B Almog}
\author{Y Dubi}
\author{G Deutscher}
\affiliation{Raymond and Beverly Sackler School of Physics and Astronomy, Tel-Aviv University, 69978 Tel-Aviv, Israel}
\date{\today}

\begin{abstract}
We present tunneling measurements of sub-micron metal/insulator/graphene planar tunnel junctions up to room temperature. We observe a gate independent gap, as previously observed only by low temperature STM[Y. Zhang et al., Nat. Phys. 4, 627 (2008)]. No gap appears at temperatures above $150K$, which is four times smaller than the theoretically expected $T_{c}$, from the accepted mean field model[T. O. Wehling et al. Phys. Rev. Lett. 101, 216803 (2008)]. We show that taking into account an additional vibrational effect of out-of-plane phonon soft modes the gap may disappear from the measurements at temperatures much lower than the calculated $T_{c}$.
\end{abstract}

\maketitle

Graphene~\cite{GrapheneReviewGeim} is famous for its linear dispersing electron and hole bands which touch at two nonequivalent points K and K' at the corner of the Brillouin zone, so it is gapless. The Fermi level of graphene lies at these Dirac points. Hence, it was a surprise, that instead of a linear V-shape conductance characteristic, a  $\sim70meV$ gap, in the tunneling spectra in STM experiments was observed~\cite{GrapheneSTMGapNatPhys,GrapheneSTMGapAPL} at low temperatures. The characteristics were symmetric and gate independent gap around zero bias voltage, which means it opens up at the Fermi level rather than at the Dirac point. The gap was interpreted~\cite{PhononGapTheory} as due to an electron-phonon interaction that couples Dirac electrons at the K point ($\pi$ band) with nearly free electrons at the $\Gamma$ point ($\sigma$ band), where the gap opens up. 

We present tunneling measurements of metal / Al-oxide / graphene sub-micron planar junctions up to room temperature. The broad temperature range is made possible by the planar geometry. Our data presents the same tunneling gap as seen so far only using STM at low temperatures. Unexpectedly we find that it disappears at $\sim150K$, a temperature 4 times smaller than expected from the accepted model~\cite{PhononGapTheory}. We suggest an addition to this model, whereby we take into account the vibrational effect of out of plane soft mode phonons. This may cause the gap to disappear at a temperature much lower than the mean field $T_{c}$.

\begin{figure}
  \begin{center}
  	\includegraphics[height=0.6\linewidth, width=1\linewidth,angle=0]{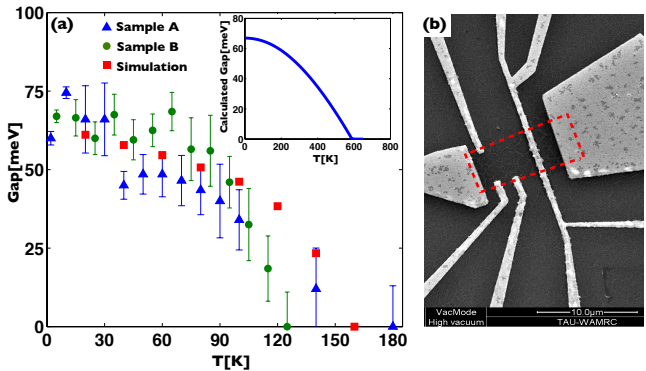}
      	\caption{a) Extracted gap value as a function of temperature, as described in the text. Green and blue symbols are the values extracted from the data. The red squares are extracted from the simulated data, Fig.~\ref{fig:ModelCalc}. Inset - Calculation of the order parameter as a function of temperature, showing that the calculated gap closes at 590K. The gap value is $\frac{\Delta^{2}}{\epsilon_{0}}$. b) Scanning Electron Microscope image of sample A. The red dashed lines outline the flake.}
	 \label{fig:GapVsTSEM}
  \end{center}
\end{figure}

We fabricated sub-micron metal / insulator / graphene junctions, as is shown in Fig. \ref{fig:GapVsTSEM}. The flake is outlined. We used commercially available single layer flakes~\cite{GrapheneInd} exfoliated on highly doped Si substrate with a 300nm oxide. The flakes have a typical mobility of a few thousand $cm^{2} / V s$. The patterns were defined using standard e-beam lithography. In the first step metallic Ti/Au contacts were evaporated using e-gun. In the second step we evaporated $20\AA$ of Al in $O_{2}$ gas at partial pressure of $10^{-4} Torr$ with the substrate cooled to 77K. The oxide layer covers the whole region of the flake. In the final step we evaporate an additional electrode, on top of the oxide and across the flake, this serves as the top electrode of the junction. We fabricated junctions with Pb electrodes (sample A) and with Ti/Au electrodes (sample B) and obtained similar results with both. We performed DC differential conductance measurements using Keithley 6221+2182A DeltaMode.

\begin{figure}
  \begin{center}
  	\includegraphics[height=.25\textheight, width=1\linewidth,angle=0]{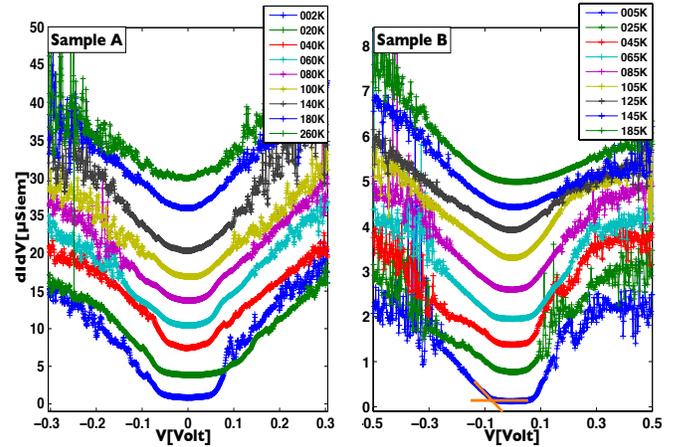}
      	\caption{Differential conductance as function of temperature. The data is shifted up for clarity. The intersection of the two orange lines, drawn on samples B's 5K curve,  exemplify the extracted gap value.}
	 \label{fig:IVTemps}
  \end{center}
\end{figure}

The low temperature differential conductance spectra of  the junctions, as shown in Fig.~\ref{fig:IVTemps}, shows a gap of $\sim70meV$. The gap is constant for all gate voltages, measured from -50V to 50V. The gap is characterized by a relatively constant conductance at low bias voltage followed by a rapid rise in the conductance. We chose the bias voltage of this intersection point to represent the gap value, as illustrated by the two orange lines, drawn on sample B's 5K curve in Fig.~\ref{fig:IVTemps}. Fig.~\ref{fig:GapVsTSEM} shows the extracted gap closes at $\sim150K$.


The low temperature gap is the same, in size, as  observed by STM~\cite{GrapheneSTMGapNatPhys,GrapheneSTMGapAPL} , which shows that it can be probed using sub-micron planar junctions. The experimental~\cite{GrapheneSTMGapNatPhys} and theoretical~\cite{PhononGapTheory} interpretation is that this is a phonon assisted tunneling gap, where the phonon is an intrinsic out of plane mode with energy of $67meV$ which hybridizes the $\pi$ (Dirac electrons) and $\sigma$ (nearly free electrons) bands. As a consequence a gap in the $\sigma$ band (nearly free electrons) opens up and a rigid shift of the energies in the $\pi$ band (Dirac electrons) occurs.

To calculate the temperature dependence of the gap we take the model interaction Hamiltonian given by T. Wehling \textit{et. al.}~\cite{PhononGapTheory} 

\begin{equation}
H_{int}=\lambda \sum_{q,k}(d_{k+q}^{\dagger} c_{q}+c_{k+q}^{\dagger}d_{q})(a_{k}+a_{-k}^{\dagger}).
\end{equation}

Where $d_{q}$ is the annihilation operator of an electron in the $\sigma$ band, in this band the electrons are nearly free electrons and may be approximated by a flat band with energy $\epsilon_{0}=3.3eV$. $c_{q}$ is the annihilation operator of an electron in the $\pi$ band(Dirac electrons), $a_{k}$ is the annihilation operator of an out of plane phonon with momentum k and $\lambda$ is the electron-phonon coupling strength. We apply a BCS-like mean field approach and define an \textit{order parameter}, $\Delta_{k}=\sum_{k'}V_{k,k'}<c_{k'}^{\dagger}d_{k'}>$. The self energies of the problem are then $E_{\pm}=\frac{\epsilon_{k}+\epsilon_{0}}{2} \pm \sqrt{(\frac{\epsilon_{k}-\epsilon_{0}}{2})^{2} + \Delta_{k}^2}$.

The excitation spectrum has a gap of size $\frac{\Delta_{k}^{2}}{\epsilon_{0}}$ in the nearly free electron band, and a rigid shift of $-\frac{\Delta_{k}^{2}}{\epsilon_{0}}$ in the energies of the Dirac electrons, in exact agreement with the results of T. Wehling \textit{et. al.}~\cite{PhononGapTheory}. We can now proceed and write the gap equation:

\begin{equation}
\Delta_{k}=\sum_{k'}V_{k,k'}\frac{\Delta_{k'}}{\sqrt{(\frac{\epsilon_{k}-\epsilon_{0}}{2})^{2} + \Delta_{k}^2}}(f_{d}(E_{-})-f_{d}(E_{+}))
\end{equation}

Where $f_{d}$ is the Fermi function. If we take $\Delta_{k,k'}=\Delta$ and $V_{k,k'}=V$, such that the gap is momentum independent, we can numerically solve the equation, where we used the experimental gap value $\frac{\Delta_{(T=0)}^{2}}{\epsilon_{0}}=67meV$. The calculated temperature dependance of the gap $\frac{\Delta^{2}}{\epsilon_{0}}$ is shown in the inset of Fig.~\ref{fig:GapVsTSEM}. The gap closes at 590K. In the calculation we took a momentum independent interaction, yet taking some form of a momentum dependent $V_{k,k'}$ will correspond to stronger coupling and only raise the calculated Tc. In contradiction to the calculated $T_{c}$  of $590K$, the observed gap disappears from the measurements at about 150K. Thermal smearing will come into effect only at temperatures above 400K and cannot explain the disappearance of the gap at about 150K.



To see how this difference can be reconciled we must first understand why, when tunneling to the $\pi$ band, the measurement is sensitive to a gap in the $\sigma$ band. Following T. Wehling \textit{et. al.}~\cite{PhononGapTheory}, the tunneling conductance is proportional to the density of states of each band multiplied by the density of the electronic wavefunction at the position of the counter electrode $dI/dV \sim |\Psi_{\Gamma}|^{2} N_{\Gamma}(E) + |\Psi_{K}|^{2} N_{K}(E)$. On the graphene sample $|\Psi_{\Gamma}|^{2}$ is negligible compared with $|\Psi_{K}|^{2}$. But $|\Psi_{K}|^{2}$ decays away from the surface, while $|\Psi_{\Gamma}|^{2}$ increases up to a distance of $\sim 1\AA$ from the surface and beyond that exponentially decays into the vacuum. The two densities decay on different length scales, and for distances larger than $1\AA$ their ratio is $ \frac{|\Psi_{\Gamma}|^{2}}{|\Psi_{K}|^{2}}=e^{1.7 h}$, where h is the distance in angstroms between the graphene sheet and the electrode. Accordingly when the electrode is far away from the graphene sheet the tunneling conductance is mostly into the $\sigma$ band, but close to the sample, it is into the $\pi$ band. At the typical distances used in STM and planar junction experiments, tunneling is mostly into the $\sigma$ band, which is why a gap is observed.

The idea behind our suggested explanation is that at high enough temperature the excited out of plane phonon soft modes will give rise to vibrations large enough to effectively bring the graphene sheet close enough to the electrode, such that tunneling will be directly to the $\pi$ band and the gap will not appear in the data, despite it's existence. The amplitude $A(\omega)$ of a phonon of frequency $\omega$ is $\sqrt{\frac{\hbar}{\rho V \omega}}$, where $\rho$ is the density and $V$ is the volume of the sample. Soft mode phonons, at low energies, have a high density of states. Also the existence of such modes was considered~\cite{GeimRTSiO2,FluxPhonGeimExp,FluxPhonPRBTheor} as the mechanism behind the additional unexpected rise in resistivity in graphene, above 150K on $SiO_{2}$~\cite{GeimRTSiO2}, and above 10K in suspended graphene~\cite{FluxPhonGeimExp}. This suggests that the contribution of these low energy excitations cannot be overlooked.

M. Mohr et al.~\cite{PhononsInGraphene} showed that graphene has an out of plane soft mode band (ZA). To evaluate the effect of the vibrations we calculate $\sqrt{<A^{2}(\omega)>}$ as a function of temperature, 

\begin{equation}
<A^{2}(\omega)>=\int A^{2}(\omega) f_{b}(\omega,T) D(\omega) d\omega.
\end{equation}

Where $f_{b}$ is the Bose function and $D(\omega)$ is the phonon density of states. The density of graphene is $\rho=7.6~10^{-7} \frac{Kg}{m^{2}}$. We took the dispersion relation as $k(\omega)=0.13\omega^{0.58}$, by fitting the out of plane acoustic phonons band (ZA)\cite{PhononsInGraphene}, where $\omega$ is in $meV$ and $k$ is in $\frac{1}{\AA}$. The fit is suitable for energies up to $30meV$, beyond these energies $f_{b}(\omega,T) D(\omega)$ is practically zero. Since the integrand diverges at zero energy we must use a cutoff parameter $\omega_{cutoff}$.

We now plug the contribution of the vibrations into the conductance. The sum of the densities $|\Psi_{\Gamma}|^{2} + |\Psi_{K}|^{2}=1$ and the ratio, away from the surface, is $\frac{|\Psi_{\Gamma}|^{2}}{|\Psi_{K}|^{2}}=e^{1.7 h}$ so the fraction of the $\pi$ band electrons is,

\begin{equation}
|\Psi_{K}|^{2}=\frac{1}{1+e^{1.7(h_{0}-\sqrt{<A^{2}>})}}
\end{equation}

\begin{figure}
  \begin{center}
  	\includegraphics[height=0.5\linewidth, width=1\linewidth,angle=0]{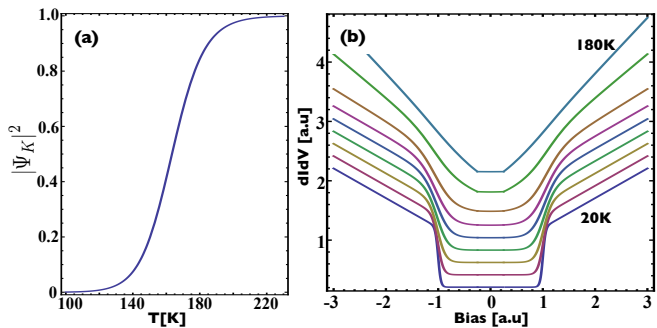}
      	\caption{a) The amplitude of the wavefunction of the Dirac electrons at the position of the tunnel electrode plotted as function of temperature. b) Simulated data realizing the transition from a $\sigma$ band density of states to a $\pi$ band density of states. Simulated data is equally spaced in temperature from 20K to 180K and shifted up for clarity. The lowest temperature (20K) simulated plot corresponds to the function we used to represent the $\sigma$ band and the highest temperature (180K) simulated plot corresponds to the function we used to represent the $\pi$ band.}
	 \label{fig:ModelCalc}
  \end{center}
\end{figure}

Where $h_{0}$ is the distance in angstroms between the electrode and the graphene sheet at zero temperature. The tunneling probability to the $\pi$ band will substantially increase as the vibrations approach the value of $h_{0}$ (2nm). The dependence of $|\Psi_{K}|^{2}$ on the temperature, shown in panel 'a' of Fig. \ref{fig:ModelCalc}, shows there is a defined transition temperature, where the contribution of $|\Psi_{\Gamma}|^{2}$ is diminished and $|\Psi_{K}|^{2}$ becomes dominant. In this calculation we used $\omega_{cutoff}=2.2\times10^{-5} meV$. Since the ratio $\frac{|\Psi_{\Gamma}|^{2}}{|\Psi_{K}|^{2}}=e^{1.7 h}$ holds only away from the surface, and closer to the surface the ratio decays rapidly to zero, the transition beyond $140K$ will be more rapid in temperature than illustrated in panel 'a' of Fig.\ref{fig:ModelCalc}. We extracted the gap from the simulated data, the results are plotted in Fig.~\ref{fig:GapVsTSEM}.

Since the oxide thickness ($h_{0}=2nm$) is experimentally known, the only parameter in the calculation is the cutoff frequency $\omega_{cutoff}$, which was chosen to give a transition temperature similar to the one observed experimentally. It corresponds to a cutoff wavelength $\lambda_{cutoff} = 2.4 \mu m$, which is a reasonable number slightly smaller than the dimensions of the physical flake. The actual flake size doesn't define $\lambda_{cutoff} $ since all the measurements show a similar transition temperature. Our calculation is based on the graphene modes\cite{PhononsInGraphene} as measured in graphite. Yet in our experiment the flake is partially pinned to the $SiO_{2}$ substrate and has the Al-oxide on top. While in both cases (graphite and "oxide sandwich") the graphene is weakly coupled to the top and bottom layers and probably has a very similar phonon spectrum, the different environments may alter the spectrum of the vibrations. More importantly a close inspection of the SEM image, Fig. \ref{fig:GapVsTSEM}, shows a non-uniform and possibly porous structure on the scale of tens of nanometers which belongs to the Al-oxide layer. Such a non-uniformity means that at a fixed temperature the vibrations will be different at different positions, so in the thinner regions of the Al-oxide layer the transition will occur when the amplitude of the vibrations are substantially smaller than $h_{0}$ and that the thicker regions may contribute less to the transition.  Therefore the transition temperature is actually a result of some averaging over the junction region. In a STM measurement, where there is no oxide separating the graphene and top electrode, so the transition temperature may be lower.

In summary we have shown the ability to fabricate planar tunnel junctions to graphene that are sensitive to the tunneling density of states. So far this has only been achieved using STM. Our technique opens up the possibility to tunnel into graphene in sub-micron sized devices. We measured the phonon assisted tunneling gap over a broad temperature range. Unexpectedly we found that the gap disappears at $\sim$150K, although the $T_{c}$ calculated from the accepted mean field model is 4 times larger. We suggest that the early disappearance of the gap may be due to the effect of out of plane soft mode phonons, whose thermal excitations bring the wavefunction of the $\pi$ band electrons, which have no gap, close to the counter electrode.

\end{document}